\documentclass[runningheads]{svmult}

\usepackage{makeidx}   % allows index generation
\usepackage{graphicx}  % standard LaTeX graphics tool
\usepackage{subeqnar}  % subnumbers individual equations
\usepackage{multicol}  % used for the two-column index
\usepackage{physprbb}  % modified textarea for proceedings,
\makeindex             % used for the subject index
\begin{document}
%
%\title*{Chaos, Chaotic Phase Mixing, and Galaxy Evolution}
\title*{Chaos and Chaotic Phase Mixing in Galaxy Evolution and Charged
Particle Beams}
%
%
%\toctitle{Chaos, Chaotic Phase Mixing, and Galaxy Evolution}
\toctitle{Chaos and Chaotic Phase Mixing in Galaxy Evolution and Charged
Particle Beams}
%Focusing of a Parallel Beam to Form a Point
%
%
\titlerunning{Chaos in Galaxy Evolution and
Charged Particle Beams}
% allows abbreviation of title, if the full title is too long
% to fit in the running head
%
\author{Henry E. Kandrup}
\authorrunning{Henry E. Kandrup}
\institute{University of Florida, Gainesville, FL 32611, USA}

\maketitle              % typesets the title of the contribution

\begin{abstract}
This paper discusses three new issues that necessarily arise in realistic
attempts to apply nonlinear dynamics to galaxy evolution, namely: (i) the 
meaning of chaos in many-body systems, (ii) the time-dependence of the bulk 
potential, which can trigger intervals of {\em transient chaos}, and 
(iii) the self-consistent
nature of any bulk chaos, which is generated by the bodies themselves, rather
than imposed externally. Simulations and theory both suggest strongly
that the physical processes associated with galactic evolution should also 
act in nonneutral plasmas and charged particle beams. This in turn suggests 
the possibility of testing this physics in real laboratory experiments, an 
undertaking currently underway.
%The abstract\index{abstract} is optional. If present it should summarize
%the contents of the paper in at least 70 and at most 150 words; neither
%too long nor too short but to the point!
\end{abstract}

\section{Introduction}
As recently as 1990, most galactic dynamicists ignored completely the possible
role of chaos in galaxies. However, the past decade has seen a growing 
recognition that chaos can be important in determining the structure of real 
galaxies. 
%which, in many cases, are less symmetric than had originally been assumed. 
Still, much recent work involving chaos in galactic astronomy has 
been simplistic in that it has involved naive applications of 
standard techniques from nonlinear dynamics
developed to analyse two- and three-degree-of-freedom 
time-independent Hamiltonian systems. The object here is to discuss some of 
the additional complications which arise if nonlinear dynamics is to be 
applied to real galaxies, which are {\em many-body systems} comprised of a 
large number of interacting masses and characterised by a {\em 
self-consistently determined bulk potential}
which, during their most interesting phases, can be 
{\em strongly time-dependent.}

\section{Transient chaos and collisionless relaxation}
\subsection{Transient chaos induced by parametric resonance}
It is well known to nonlinear dynamicists that the introduction of an
oscillatory time-dependence into even an otherwise integrable potential can 
trigger an interval of {\em transient chaos}, during which many orbits 
exhibit an exponentially sensitive dependence on initial conditions. 
Physically, this transient chaos arises from a resonance overlap between the 
frequencies ${\sim}{\;}{\Omega}$ of the unperturbed orbits and the frequency 
or frequencies ${\sim}{\;}{\omega}$ of the time-dependent perturbation.

In the past, the possible effects of such chaos have been considered for 
both nonneutral plasmas~\cite{Stras} and charged particle beams~\cite{Gluck}. 
More recently, such transient chaos has also begun to be considered in the 
context of galactic astronomy~\cite{KVS}.
That work has shown that, for large fractional amplitudes, $>0.1$ or so, 
this resonance can be very broad, triggering significant amounts of chaos for 
$10^{-1}\;{\le}\;{\omega}/{\Omega}\;{\le}\;10$; and that the existence of the 
phenomenon is robust, comparatively insensitive to details. It will, for 
example, persist if one allows for damped oscillations and/or modest drifts 
in frequency (generated, {\em e.g.,} by making ${\omega}$ a random variable
sampling an Ornstein-Uhlenbeck colored noise process).

The breadth of the resonance and the insensitivity to details suggest that
transient chaos could well prove common, if not generic, in violent 
relaxation~\cite{LB67}, the collective process whereby a (nearly)
collisionless galaxy or galactic halo evolves towards an equilibrium or 
near-equilibrium state. Violent relaxation typically involves damped 
oscillations triggered, {\em e.g.,} by interactions with another galaxy 
or, in the early Universe, by the cosmological details preceding galaxy 
formation. However, when considering collective effects there is only 
one natural time scale, the dynamical time 
$t_{D}{\;}{\sim}{\;}1/\sqrt{G{\rho}}$, with ${\rho}$ a typical mass density, 
which determines both the characteristic orbital time scale and (at least 
initially) the oscillation time scale. The exact numerical values of these
time scales will involve numerical coefficients which will in general be 
unequal and vary as a function of location within the galaxy. If, however,
one need only demand that the oscillation and orbital time scales agree to
within an order of magnitude, it would seem likely that this resonance could
trigger transient chaos through large parts of the galaxy.
In real galaxies the oscillations will presumably damp and the frequencies
drift as the density changes and, presumably, power cascades from longer
to shorter scales. 
To the extent, however, that the details are unimportant such variations should
not obviate the basic effect.

\subsection{Chaotic phase mixing and collisionless relaxation}
But why might such transient chaos prove important in galactic evolution?
Detailed numerical simulations indicate that violent relaxation can be a very 
rapid and efficient process, but simple models involving regular orbits, 
such as Lynden-Bell's~\cite{LB67} balls rolling in a pig-trough, do not
approach an equilibrium  nearly fast enough~\cite{Kan99}. The important point,
however, is that allowing for the effects of chaos can in principle 
dramatically accelerate both the speed and efficacy of violent relaxation.
An initially localised ensemble of regular orbits evolved into the future
in a time-independent potential will begin by diverging as a power law in time
and, only after a very long period, slowly evolve towards a time-averaged
equilibrium,
{\em i.e.,} a uniform population of the {\em KAM} tori to which it is 
restricted. By contrast, a corresponding ensemble of chaotic orbits will begin
by diverging exponentially at a rate that is comparable to a typical value of
the largest finite time Lyapunov exponent for the orbits in the ensemble; and
then converge exponentially towards an equilibrium or near-equilibrium state
at a somewhat smaller, but still comparable, rate. The exponential character
of this {\em chaotic phase mixing} means that the time scale associated with
this process is typically far shorter than the time scale associated with
{\em regular phase mixing}~\cite{KM94,MABK,MV}.

It is evident that chaotic phase mixing in a time-independent Hamiltonian 
system can trigger a very rapid approach towards an equilibrium, but this
does not necessarily `explain' violent relaxation. If, {\em e.g.,} most of
the orbits in the system are regular, it would seem unlikely that chaotic 
phase mixing could be sufficiently ubiquitous to explain an approach towards a
(near-)equilibrium for the galaxy as a whole. Indeed, one would expect that, 
for a galaxy that is in or near equilibrium the relative measure of chaotic 
orbits should be comparatively small: If the galaxy exhibits nontrivial 
structures like a bar or a cusp, the types of structures which one has come
to associate with chaos, one would also expect large numbers of regular
orbits must be present to serve as a `skeleton' to support that 
structure~\cite{B78}. Moreover, it is evident that, although chaotic mixing
in a time-independent potential can be very efficient in mixing orbits on a
constant energy surface, the energy of each particle remains conserved, so
that there can be no mixing in energies. The extent to which chaotic phase 
mixing in a time-dependent potential will trigger an efficient shuffling of 
energies is not completely clear.

The important point, then, is that chaotic phase mixing associated with 
transient chaos in a time-dependent potential is likely to explain these 
remaining
lacunae. At least for large amplitude perturbations, (say) $10\%$ or more, 
this 
parametric resonance can trigger a huge increase in the relative abundance
of chaotic orbits so that, for pulsation frequencies near the middle of the
resonance, virtually all the orbits exhibit substantial exponential 
sensitivity. Moreover, given that this chaos involves a resonant coupling,
it tends typically to cause a substantial shuffling of energies: those
frequencies which are most apt to trigger lots of chaos are also apt to 
induce the largest shuffling of energies.
\begin{figure}[b]
\begin{center}
\includegraphics[width=.7\textwidth]{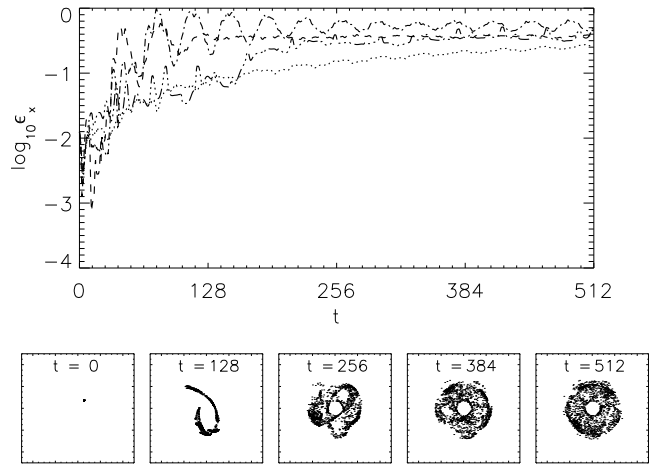}
\end{center}
\caption[]{(top) The emittance ${\epsilon}_{x}$ for an ensemble of initial 
conditions evolved in an integrable Plummer potential subjected to damped
oscillations with four different frequencies. (bottom) $x$-$y$ scatter plots 
corresponding
to the uppermost curve.}
\label{fig1.eps}
\end{figure}

Still it should be noted that one {\em can} get a `near-complete' shuffling of
orbits on different constant energy surfaces even if the orbital energies are 
not especially well shuffled. This, however, is not necessarily a problem. 
Simulations of systems exhibiting efficient collective relaxation do not 
necessarily involve masses which completely `forget' their initial conditions. 
Rather, comparatively efficient and complete violent relaxation is completely 
consistent with an evolution in which masses `remember' (at least partially)
their initial
binding energies, {\em i.e.,} in which masses that start with comparatively
large (small) binding energies end up with comparatively large (small) binding
energies~\cite{ZQ}.

That it may be possible to achieve efficient chaotic phase mixing in an
oscillating galactic potential while still relaxing towards a nearly integrable
state within $10t_{D}$ or so is illustrated in Fig.~\ref{fig1.eps}. This
Figure was generated from orbits evolved in a time-dependent potential of the 
form
\begin{equation}
V(x,y,z,t)=-{m(t)\over (1+x^{2}+y^{2}+z^{2})^{1/2}}, \qquad
m(t)=1+{\delta}m{\sin {\omega}t\over (t_{0}+t)^{p}},
\end{equation}
with ${\delta}m=0.5$, $t_{0}=100$ and $p=2$, which represents a 
galaxy damping towards an integrable Plummer sphere. 
The four curves in the top panel exhibit the $x$-component of the phase 
space {\em emittance},
${\epsilon}_{i}=({\langle}r_{i}^{2}{\rangle}{\langle}v_{i}^{2}{\rangle}-
{\langle}r_{i}v_{i}{\rangle}^{2})^{1/2}$ ($i=x,y,z$),
all computed for the same localised ensemble of initial conditions, but 
allowing
for four different frequencies ${\omega}$. The curves exhibit considerable
structure but, at least for early times, the overall evolution is exponential.
The bottom panels exhibit the
$x$ and $y$ coordinates at five different times for the ensemble represented
by the uppermost of the four curves. Here $t_{D}{\;}{\sim}{\;}20$, so that
$t=256$ corresponds to roughly $12t_{D}$.  

Intuitively, one might expect that strong oscillations, which trigger the
largest finite time Lyapunov exponents and the largest number of chaotic
orbits, would yield the fastest chaotic phase mixing and, hence, the
most rapid and most complete violent relaxation. A time-dependence with a
weaker oscillatory component, {\em e.g.,} a time-dependence corresponding to
a near-homologous collapse, might instead be expected to yield less chaos and,
hence, less efficient and less complete violent relaxation.
There is, therefore, an important need to determine
the extent to which, in real simulations of violent relaxation, many/most 
of the orbits (or phase elements) are strongly chaotic, and the degree to
which  the rate and completeness of the observed violent relaxation correlate
with the size of the largest finite time Lyapunov exponents and/or the
relative measure of chaotic orbits. Investigations of these issues are
currently underway.

\section{The role of discreteness effects}
\subsection{Microchaos and macrochaos}
The discussion in the preceding section, like most applications of nonlinear
dynamics to galactic astronomy, neglects completely discreteness effects
associated with the `true' many-body potential, assuming that masses in a
galaxy can be approximated as evolving in a smooth, albeit time-dependent, 
three-dimensional potential and that `chaos' has its usual meaning.
That this is justified is not completely obvious. 
The gravitational $N$-body problem for a large number of bodies of comparable
mass is strongly chaotic in the sense that individual orbits have large
positive Lyapunov exponent ${\chi}_{N}$ {\em even when there is absolutely no 
chaos in the continuum limit!} If, {\em e.g.,} a smooth density distribution 
corresponding to an integrable potential is sampled to generate an 
$N$-body density distribution, one finds that orbits evolved in this 
$N$-body distribution will be strongly chaotic, even for very large $N$,
despite the fact that characteristics in the smooth potential generated from 
the same initial condition are completely integrable. Moreover, there is no 
sense in which the exponential sensitivity decreases with increasing $N$:
if anything ${\chi}_{N}$ is an increasing function of $N$~\cite{HM}. In this
sense, {\em larger $N$ implies more chaos, not less!}

This situation has led some astrophysicists to question, either implicitly
or explicitly, the reliability of the entire smooth potential approximation. 
Thus, {\em e.g.,} it has been suggested \cite{Heg} that ``the approximation of 
a smooth potential is useful for studying orbits, but not for studying their 
divergence.''  This is of course a problematic statement in that the 
distinction between exponential and power law divergence, emblematic of the 
differences between regular and chaotic behaviour, lies at the heart of 
applications of nonlinear dynamics to galactic dynamics. If the Lyapunov 
exponents associated with the bulk potential have nothing to do with the 
$N$-body problem, one must perforce reject completely all conventional 
applications of nonlinear dynamics to galactic astronomy.

The crucial point, then, is that there {\em does} appear to be a 
well-defined continuum limit, even at the level of individual 
orbits~\cite{KS01,SK02,KS03}. Suppose that a smooth density distribution,
corresponding either to an integrable potential or to a potential admitting
large measures of regular orbits, is sampled to generate a fixed, {\em i.e.}
frozen in time, $N$-body density distribution, and that the trajectories
of test particles evolved in this frozen distribution are compared with
smooth potential characteristics with the same initial conditions. In this
case, there is a precise sense in which, as $N$ increases, the frozen-$N$
trajectories converge towards the smooth potential characterstics. Both 
visually and in terms of the {\em complexity}~\cite{KEB} of their Fourier 
spectra, the frozen-$N$ trajectories come to more closely resemble the smooth
potential characteristics; and, {\em viewed mesoscopically}, the frozen-$N$
and smooth potential orbits remain closer in phase space for progressively
longer times. In particular, a frozen-$N$ orbit corresponding to an integrable
characteristic will have a large Lyapunov exponent ${\chi}_{N}$ even if,
visually, it is essentially indistinguishable from the regular characteristic!

But how can this be? The key recognition here is that two `types' of chaos 
can be present in the $N$-body problem, characterised by {\em two different 
sets of Lyapunov exponents associated with physics on different scales.}
Close encounters between particles trigger {\em microchaos}, a generic feature
of the $N$-body problem, which leads to large positive Lyapunov exponents
${\chi}_{N}$. If, however, the bulk smooth potential is chaotic, one will also 
observe {\em macrochaos}, which is again characterised by positive, albeit 
typically much smaller, Lyapunov exponents ${\chi}_{S}$. Suppose, for example, 
that one compares the evolution of two nearby chaotic initial conditions in a 
single frozen-$N$ background or the same chaotic initial condition evolved in 
two different frozen-$N$ realisations of the same bulk density. In this case,
one typically observes a three-stage evolution, namely: (1) a rapid 
exponential divergence at a rate ${\chi}_{N}$ set by the true Lyapunov 
exponents associated with the $N$-body problem, which persists until the 
separation becomes large compared with a typical interparticle spacing; 
followed by (2) a slower exponential divergence at a rate comparable to the 
(typically much smaller) smooth potential Lyapunov exponent ${\chi}_{S}$, 
which persists until the separation becomes macroscopic; followed by (3) a 
power law divergence on a time scale ${\propto}{\;}(\ln N)t_{D}$.
For regular initial conditions, the second stage is absent and the
time scale for the third stage scales instead as $N^{1/2}t_{D}$. 

Microchaos becomes stronger as $N$ increases in the sense that the value of
${\chi}_{N}$ increases with increasing $N$~\cite{Pogo}. Despite this, however,
it becomes progressively less important macroscopically in that the {\em range}
of the chaos, {\em i.e.,} the scale on which the microchaos-driven exponential 
divergence of nearby orbits terminates, decreases with increasing $N$. 
In the limit $N\to\infty$ microchaos will become completely irrelevant but,
for finite $N$, it does have an effect, at least on sufficiently short scales;
and it is possible from an $N$-body simulation to extract estimates of both
${\chi}_{N}$ and the typically much smaller ${\chi}_{S}$~\cite{KS03}.

\subsection{Modeling discreteness effects as friction and noise}
It has been long recognised that, for sufficiently small $N$ and/or over
sufficiently long times, discreteness effects will not be completely 
negligible. 
Systems like galaxies are `nearly collisionless' in the sense that the stars
interact primarily via collective macroscopic forces associated with the
bulk density distribution; but, at least in principle, if one waits long
enough discreteness effects should have an appreciable effect.

Astronomers are accustomed to modeling discreteness effects in the context of a
Fokker-Planck description analogous to that formulated originally in the
context of plasma physics~\cite{RMJ}. However, it is not completely clear
to what extent this is really justified. The conventional Fokker-Planck 
description was formulated originally to extract statistical properties of 
orbit ensembles and distribution functions over long time scales, assuming 
implicitly that the bulk potential is regular. To what extent, then, can 
Langevin realisations of a Fokker-Planck equation yield reliable information 
about individual orbits over comparatively short time scales, particularly if 
the orbits are chaotic? 

Analyses of flows in frozen-$N$ systems indicate~\cite{SK02} that, at the 
level of both orbit ensembles and individual orbits, discreteness effects 
can in fact be modeled {\em extremely} well by Gaussian white noise in the 
context of a Fokker-Planck description, allowing for a dimensionless diffusion 
constant $D{\;}{\propto}{\;}1/N$, consistent with the predicted scaling 
$D{\;}{\propto}{\;}\ln {\Lambda}/N$, with ${\Lambda}$ the so-called Coulomb 
logarithm~\cite{RMJ}. 
For localised ensembles of initial conditions corresponding to both 
regular and chaotic orbits, phase mixing in frozen-$N$ systems and phase 
mixing in smooth potentials perturbed by Gaussian white noise yield virtually
identical behaviour, both in terms of the evolution of various phase space
moments such as the emittance and the rate at which individual orbits in the 
ensemble exhibit
nontrivial `transitions', {\em e.g.,} passing through some {\em entropy 
barrier} from one phase space region to another. And similarly, a comparison 
of frozen-$N$ orbits and noisy smooth potential orbits with the same initial
condition reveals that their Fourier spectra typically exhibit comparable 
complexities. 
Gaussian white noise is even successful in mimicking some of the effects of
microchaos.
If, {\em e.g.,} one tracks the divergence of two noisy orbits with the same 
chaotic initial condition evolved in a smooth potential, one observes the same 
three-stage evolution as for a pair of frozen-$N$ orbits evolved in two
different frozen-$N$ potentials.

\begin{figure}[b]
\begin{center}
\includegraphics[width=.7\textwidth]{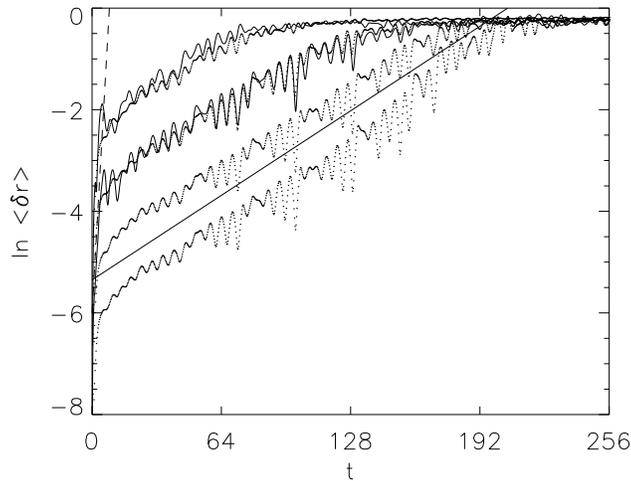}
\end{center}
\caption[]{The mean spatial separation between the same initial conditions
evolved in two different frozen-$N$ backgrounds (solid curves) and different
noisy orbits evolved in the smooth potential from the same initial condition
(dots). The solid line has a slope
$0.022$, equal to the mean value of the smooth potential Lyapunov exponent
${\chi}_{S}$. The dashed curve has a slope $0.75$, equal to the mean value of 
the $N$-body Lyapunov exponent ${\chi}_{N}$.}

\label{fig2.eps}
\end{figure}

An example of this agreement is illustrated in Fig.~\ref{fig2.eps}, which 
exhibits
data generated by averaging over $100$ pairs of orbits evolved in frozen-$N$
density distributions which correspond in the continuum limit to a triaxial
homogeneous ellipsoid with axis ratios $1.95:1.50:1.05$, perturbed by a 
spherically symmetric central mass 
spikes (`black hole'). The top two solid curves represent (from top
to bottom) results for $N=10^{4.5}$ and $N=10^{5.5}$. The four dotted curves
represent analogous results derived for pairs of noisy orbits evolved from the 
same initial conditions in the smooth potential with (from top to bottom)
diffusion constant $D=10^{-4}$, $10^{-5}$, $10^{-6}$, and $10^{-7}$. The
near-coincidence of the top two solid and dotted curves indicates that
discreteness effects for $N=10^{p+1/2}$ are well-mimicked by Gaussian white
noise with $D=10^{-p}$.

Such striking agreement 
suggests strongly that investigations of how orbits in smooth
potentials are impacted by the introduction of friction and noise can provide
important insights into the role of graininess in real galaxies. 
It is customary to assert that, in a system as large as a galaxy, discreteness 
effects reflecting close encounters between stars are unimportant because 
the relaxation time $t_{R}$ on which they can induce appreciable changes in 
quantities like the energy is orders of magnitude longer than the age of the
Universe~\cite{BT}.
This is likely to be true if the galaxy is an exact equilibrium, especially
an equilibrium characterised by an integrable potential.
However, the assertion is suspect if (as must usually be the case) the system 
is only `close to' an equilibrium or near-equilibrium, especially if the bulk
potential is characterised by a phase space admitting a complex coexistence
of regular and chaotic orbits.

Over the past decade, analyses of flows in time-independent Hamiltonian 
systems have revealed that even very weak perturbations, idealised
as friction and white noise corresponding to 
$t_{R}{\;}{\sim}{\;}10^{6}-10^{9}t_{D}$ and, hence, 
$D{\;}{\sim}{\;}10^{-6}-10^{-9}$, can have significant effects within a time
as short as $100t_{D}$ or less by facilitating phase space diffusion through
cantori or along the Arnold web~\cite{HKM,SiopK,Nov}.
The basic point is that the motions of chaotic orbits in a complex potential 
can be constrained significantly by topological obstructions like cantori
or the Arnold web which, albeit not completely preventing motions from one
phase space region to another, serve as an {\em entropy barrier} to impede
such motions. 
In many respects, the physical picture is similar to the elementary problem 
of effusion of gas through a tiny hole in a wall. There is nothing in 
principle to prevent a gas molecule from passing through the hole and, hence,
escaping from the region to which it is originally confined; but, if the hole
is very small, the time scale associated with this effusion can be extremely
long.

In the same sense, and for much the same reason, chaotic orbits trapped in
one phase space region may, in the absence of perturbations, remain stuck in
that region for a very long time. However, subjecting the orbits to noise will 
`wiggle' them in such a fashion as to increase the rate at which they pass 
through the entropy barrier, thus accelerating phase space transport. 
Numerical simulations indicate that, in at least some cases, this escape 
process can be well approximated by a Poisson process, with the number of
nonescapers decreasing exponentially at a rate ${\Lambda}$ that is determined
by the perturbation~\cite{PK,KPS}. 
This effect appears to result from a resonant coupling between the orbits and
the noise. White noise is characterised by a flat power spectrum and,
as such, will couple to more or less anything. If, however, the noise is made
coloured, {\em i.e.,} if instantaneous kicks are replaced by impulses of finite
duration, the high frequency power is reduced; and, if the autocorrelation
time becomes sufficiently long that there is little power at frequencies
comparable to the orbital frequencies, the effect of the noise decreases
significantly. Significantly, it appears that, overall, the details of the 
perturbation may be largely irrelevant: additive and multiplicative Gaussian
noises tend to have comparable
effects and the presence or absence of friction does not seem to matter.
All that appears to matter are the amplitude and the autocorrelation time
upon which there is a relatively weak, roughly logarithmic, dependence.

But what does all this imply for a real galaxy?
Given that collisionless near-equilibria must be more common than true 
equilibria, it would seem quite possible that, during the early stages of
evolution, a galaxy might settle down towards a near-equilibrium, rather
than a true equilibrium, {\em e.g.,} involving what have been termed~\cite{MF}
`partially mixed' building blocks. If discreteness effects and all other
perturbative effects could be ignored, such a quasi-equilibrium might persist
without exhibiting significant changes over the age of the Universe. 
If, however, one allows for discreteness effects or, alternatively, other
perturbations reflecting, {\em e.g.,} a high density cluster environment,
the orbits could become shuffled in such a fashion as to trigger significant
changes in the phase space density and, consequently, a systematic secular
evolution~\cite{SSR}.

Such a scenario could, for example, result in the destabilisation of a bar.
Many models of bars ({\em e.g.}~\cite{paq}) incorporate `sticky'~\cite{Con} 
chaotic orbits as part of the skeleton of structure, replacing crucial regular 
orbits which can be absent near corotation and other resonances. Making these 
`sticky' orbits become unstuck could cause the bar to dissipate.
Similar effects could also cause an originally nonaxisymmetric cusp to
evolve towards a more nearly axisymmetric state. To the extent that the
triaxial Dehnen potentials are representative, one can argue that chaotic
orbits may be extremely common near the centers of early-type galaxies, but
that many of these chaotic orbits are extremely sticky~\cite{KSiop} and,
as such, could help support the nonaxisymmetric structure. Perturbations
that make these sticky orbits wildly chaotic could {\em de facto} break
the bones of the skeleton supporting the structure and trigger an evolution 
towards axisymmetry.

%\section{Issues related to self-consistency}

\section{Experimental tests of galactic dynamics}
\subsection{Similarities between galaxies and nonneutral plasmas}
Even though electrostatics and Newtonian gravity both involve $1/r^{2}$ forces,
electric neutrality implies that the physics of neutral plasmas is 
very different 
from the physics of self-gravitating systems. Viewed over time scales 
$>t_{R}$, nonneutral plasmas and charged-particle beams are also very
different from self-gravitating systems: the attractive character of 
gravity leads to phenomena like evaporation and core-collapse which cannot
arise in a beam or a plasma. If, however, one restricts attention to 
comparatively short times ${\ll}{\;}t_{R}$, much of the physics should be the 
same. Theoretical expectations, supported by numerical simulations, suggest
that it is the existence of long range order, not the sign of the interaction,
which is really important; but, to the extent that this be true, collisionless
plasmas and collisionless self-gravitating systems should be quite similar.

Typical sources of charged-particle beams configure the beams in trains of
`packets' or `bunches', as they are termed by accelerator dynamicists. 
The objective of a good high-intensity accelerator is to generate bunches 
comprised of a large total number of charges confined to a small phase space 
volume and then accelerate those bunches to very high energies while 
minimising any growth in emittance. As one example, modern photocathode-based
sources of electron beams routinely generate bunches comprised of some
$10^{10}-10^{11}$ electrons with transverse `emittance' 
${\tilde{\epsilon}}$ of a few microns. (Here 
${\tilde{\epsilon}}={\epsilon}/v_{0}$,
where $v_{0}$ is the mean axial velocity of the particle distribution.)
The energy relaxation time $t_{R}$ associated with such bunches typically
corresponds to the time required for a bunch to travel a distance ${\sim}{\;}1$
km or so which, in many cases, is much longer than any distance of experimental
interest, so that the bunches are `nearly collisionless.'

Models of equilibrium configurations of nonneutral plasmas and 
charged-particle beams confined by electromagnetic fields can
be characterised by a complex phase space quite similar to that associated
with models of elliptical galaxies and, as such, have orbits 
with very similar properties. 
For example~\cite{BS}, the so-called `thermal equilibrium model'~\cite{BR} of 
beam dynamics, which involves a self-interacting nonneutral plasma in thermal
equilibrium confined by an anisotropic harmonic oscillator potential, is
strikingly similar~\cite{KSiop} to the nonspherical generalisations of the 
Dehnen potential of galactic astronomy in terms of such properties as the
degree of `stickiness' manifested by chaotic orbits or how the relative 
measure of chaotic orbits and
the size of the largest Lyapunov exponent vary with shape.

As in galactic dynamics, questions have been raised regarding the validity
of the continuum approximation for nearly collisionless charged particle 
beams~\cite{Struck}. However,
comparatively short time integrations ($t{\;}{\ll}{\;}t_{R}$) involving 
discreteness effects and the nature of the continuum limit in nonneutral 
plasmas~\cite{KSB} yield results essentially identical to what is 
observed for gravity -- although the behaviour associated with neutral plasmas 
is {\em very} different. In particular, the macroscopic manifestations of 
phase mixing, for both regular and chaotic orbits, are indistinguishable, and 
the coexistence of microchaos and macrochaos persists unabated.

Nontrivial effects associated with a time-dependent potential have also been 
predicted for both nonneutral plasmas~\cite{Stras} and charged particle 
beams~\cite{Gluck}. Although the time-dependence that is envisioned in a
beam is typically less violent than that anticipated in violent relaxation
within a galaxy, such is not always the case. Indeed, there is compelling 
experimental
evidence that, in a beam, such a time-dependence can have the undesireable 
effect of ejecting particles from the core into an outerlying halo~\cite{Wan}.

\begin{figure}[b]
\begin{center}
\includegraphics[width=1.3\textwidth]{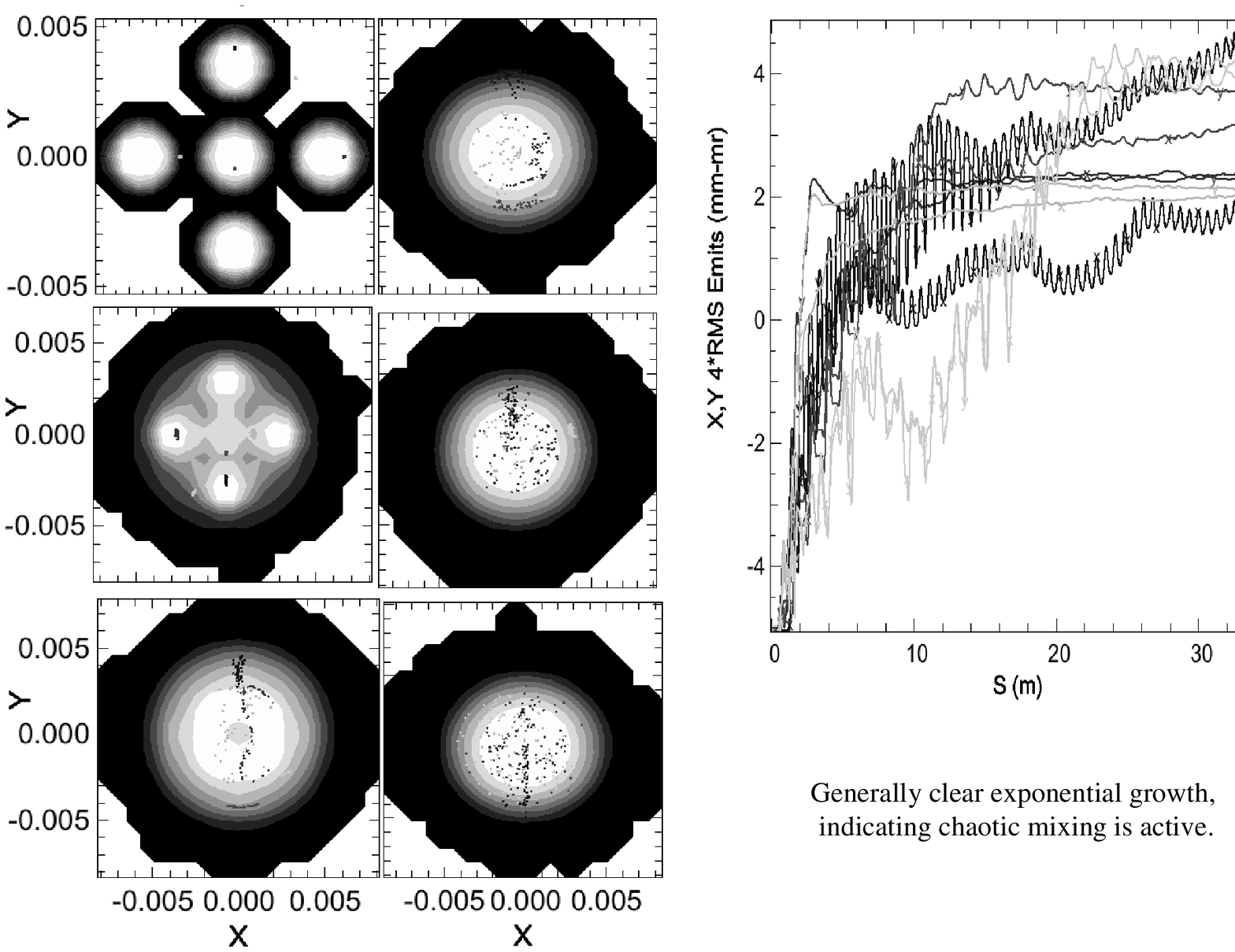}
\end{center}
%\vskip -0.4in
\caption[]{Evolution of five representative ensembles of test particles in the
five-beamlet simulation. The left hand panel shows snapshots of the ensembles
at (top-to-bottom left column) 0 m, 0.98 m, 2.88 m, and (top-to-bottom right
column) 5.24 m, 11.52 m, and 31.68 m, with the $x$- and $y$-axes labeled in
meters. The right panel shows the evolution of
the logarithm of the emittances ${\epsilon}_{x}$ and ${\epsilon}_{y}$ as a
function of distance $S(z)$ along the accelerator.
}
\label{fig3.eps}
\end{figure}

Perhaps most interesting, however, is the fact that numerical simulations
that reproduce successfully `anomalous relaxation' observed in real laboratory
experiments involving accelerator beams have shown compelling 
evidence of chaotic phase mixing. One classic example involves the propagation
of five nonrelativistic high-intensity beamlets in a periodic solenoidal 
transport channel, where self-consistent space-charge forces are extremely
important~\cite{Rei}. Ideally, these beamlets should exhibit coherent 
periodic oscillations (quite literally disappearing and reappearing) which 
might be expected to decay only on a relaxation time scale $t_{R}$ that 
corresponds to a propagation distance ${\sim}{\;}1$ km. However,
regardless of how well the beam was matched to the transport channel, the 
beamlets were seen to reappear only once, at a point ${\sim}{\;}1$ m from 
the source, disappearing completely within 2 m or so ({\em cf.} Fig.~[6.10] 
in \cite{Rei}). Their failure to reappear 
again would seem to reflect some collisionless process that, in effect, causes 
the particles to `forget' their initial conditions.

Detailed simulations using the particle-in-cell code {\em WARP}~\cite{WARP}, 
which do an extremely good job of reproducing what is actually seen, 
demonstrate seemingly unambiguously that, because of the 
time-dependent space-charge potential, a large fraction of the particles in 
the beam experience the effects of strong, possible transient, 
macrochaos~\cite{KSR,Ketal}. This is, {\em e.g.,} evident from Fig.~3, which
illustrates the evolution of representative test particles which interact
with the bulk potential but not with each other. Here the left hand panel
shows snapshots of the beam after it has travelled distances $0$ m, $0.98$ m,
$2.88$ m, $5.24$ m, $11.52$ m, and $31.68$ m, with the representative 
ensembles superimposed. The right hand panel exhibits the evolution of the
emittances ${\epsilon}_{x}$ and ${\epsilon}_{y}$ for these ensembles. It is
evident that the initially localised ensembles are diverging exponentially
so as to fill much of the accessible phase space, and that this exponential
divergence coincides with the beamlets losing their individual identities.
Also evident is the fact that the behaviour observed here is very similar
to that exhibited in Fig.~1 which, recall, was generated for orbits in a
perturbed Plummer potential exhibiting damped oscillations.

\subsection{Testing galaxy evolution with charged particle beams}

The aforementioned similarities between galactic astronomy and charged 
particle beams suggest the possibility of using accelerators as a laboratory
for astrophysics in which one can perform experimental tests of galactic 
dynamics, a possibility currently being developed by a University of Florida
-- Fermilab/Northern Illinois University -- University of Maryland 
collaboration. This collaboration, which has the dual aims of (1) obtaining 
an improved understanding of the applicability of 
nonlinear dynamics to nearly collisionless systems interacting via long
range forces and (2) using that understanding to generate more sharply focused 
bunches by minimising undesirable increases in emittance, is currently
planning concrete experiments which can, and presumably will, be performed 
on the University
of Maryland Electron Ring ({\em UMER}) currently under construction.
Here a number of obvious issues, all experimentally testable, come to mind:

How ubiquitous is chaotic phase mixing as a source of anomalous relaxation?
Older experiments with lower-intensity beams, where the space-charge forces
were comparatively unimportant, tended not to manifest extreme examples of 
anomalous relaxation. Anomalous relaxation appears more common in high
intensity beams, especially in settings where a time-dependent density
distribution generates a strongly time-dependent potential; and it is obvious
to ask whether chaos is the principal culprit.
The idea here is to identify the types of scenaria that tend generically
to yield anomalous relaxation and to determine, {\em e.g.,} whether such
scenaria tend typically to be associated with a bulk potential that 
incorporates a strong, roughly oscillatory component. Do numerical simulations
of orbits ensembles evolved in such systems exhibit evidence of chaotic
phase mixing? And do individual orbits in those ensembles exhibit strong
exponential sensitivity, associated, {\em e.g.,} with transient chaos?

Do instabilities tend to trigger transient chaos? Instabilities in 
collisionless systems can exhibit behaviour qualitatively similar to
that associated with turbulence in collision-dominated systems, but it is
well known that turbulence is a strongly chaotic phenomenon. This possibility
is especially interesting in that turbulence is another setting where 
different `types' of chaos, characterised by wildly different time scales, 
can act on different length scales.

What types of geometries, both strongly time-dependent and nearly 
time-independent, tend to yield the most efficient chaotic phase mixing 
and the largest measures of chaotic orbits? Do time-dependent evolutions
involving strongly convulsive oscillations tend generically to exhibit
especially fast relaxation? And do they tend to yield especially large
amounts of chaos, as probed by the relative measure of chaotic orbits and/or
the sizes of the largest (finite time) Lyapunov exponents?
To the extent that bulk properties of such `accelerator violent relaxation' 
correlate with the degree of chaos exhibited in the evolving beam, and that 
the degree of chaos correlates with the form of the macroscopic time 
dependence, one will have a physically well-motivated explanation of which 
sorts of scenaria would be expected to exhibit complete and efficient violent 
relaxation and which would not! 

Is it, {\em e.g.,} true that, for nearly axisymmetric configurations, prolate 
(or oblate) bunches tend generically to exhibit especially large amounts of 
chaos?  And do any such trends that are observed coincide with trends observed 
in models of galactic equilibria~\cite{KSiop}? 
Even if a beam bunch remains nearly
axisymmetric during its evolution, the acceleration mechanism can -- and in
general will -- change its shape as it passes down an accelerator in a fashion
that depends on the accelerator design. The obvious question, then, is
whether the oblate or prolate phase tends generically to be especially chaotic.

Addressing this and related issues could provide important insights as to 
{\em why} galaxies have the detailed shapes that they do, a general question
for which, at the present time, no compelling dynamical explanation exists. 
One knows, {\em e.g.}, that elliptical galaxies tend to have isophotes that
are slightly boxy or disky, and this boxiness or diskiness correlates with
such properties as the rotation rate, the steepness of the central cusp, and
the size of any deviations from axisymmetry~\cite{KB}. Must all these effects
be attributed to the detailed form of the formation scenario, or is there 
a clear dynamical explanation? Is it, {\em e.g.,} true that the observed 
deviations from perfect ellipsoidal symmetry conspire to reduce the relative 
number of chaotic orbits or to increase the numbers of certain regular orbit 
types required as a skeleton to support the observed structure?

One might also use accelerator experiments to probe the role of discrete
substructures and the extent to which they can be modeled as friction and
noise in the context of a Fokker-Planck description~\cite{BD}. If the
injection of a beam involves a large mismatch, a significant charge 
redistribution will occur, resulting in violent relaxation, `turbulent'
behaviour, and the formation of substructures (`lumps') on a variety of 
scales. To the extent that such a time-dependent evolution can be described
in a continuum approximation, one might then expect that the bulk potential
will correspond to a highly complex time-dependent phase space and that
the substructures could act as a `noisy' source of extrinsic diffusion,
facilitating both transitions between `sticky' and `wildly chaotic' behaviour
and, in some cases, transitions between regularity and chaos. 
Given the evidence ({\em cf.}~\cite{KSB}) that, at least over short times, 
discreteness 
effects act similarly for attractive and repulsive $1/r^2$ forces, such
insights could be directly related to such issues as the destabilisation of 
bars in
spirals and/or the secular evolution of nonaxisymmetric ellipticals towards 
more nearly axisymmetric states.

Do systems tend to evolve in such a fashion as to minimise the amount of
chaos? There is an intuitive expectation amongst many galactic astronomers 
({\em cf.}~\cite{Sch}) that galaxies tend to evolve towards equilibria
which incorporate few if any chaotic orbits, {\em i.e.,} that nature somehow
favors nearly-regular equilibria. It would certainly appear true that a model 
must incorporate significant numbers of regular orbits to support interesting
structures like bars and/or triaxiality, but this does not {\em a priori}
preclude the possibility of chaotic orbits also being present. Generic
time-independent three-degree-of-freedom potentials are neither completely
regular nor completely chaotic, admitting instead a complex coexistence of
regular and chaotic phase space regions. The obvious question then is: 
are galactic equilibria or near-equilibria typically well-represented by
potentials which are generic in this sense; or are they, for reasons unknown,
special in that they tend to be rather nearly regular?

\section{Conclusions}
This paper has focused on several fundamental issues that arise in attempts 
to apply nonlinear dynamics to real galaxies, many-body systems characterised
by a self-consistently determined bulk potential which, during their most
interesting phases, can be strongly time-dependent. 
As recently as a decade ago these issues would have been considered of largely 
academic, rather than practical, interest.
However, recent observational advances -- which facilitate improved high
resolution photometry of individual objects as well as statistical analyses
of large samples with varying redshift -- and improved computational resources
-- which allow unparalleled explorations of multi-scale structure --,
together with the recognition that the basic physics can be also probed in the
context of charged particle beams, imply that theoretical predictions regarding
these `academic' issues can in fact be tested observationally, numerically, 
and experimentally.
\vskip .1in
\par\noindent
Thanks to Ioannis Sideris for providing Fig.~1 and to Rami Kishek for
providing Fig.~3. Thanks also to Court Bohn for his comments on a preliminary
draft of the manuscript.
Limited financial support was provided by NSF AST-0070809.

%INDEX%%%%%%%%%%%%%%%%%%%%%%%%%%%%%%%%%%%%%%%%%%%%%%%%%%%%%%%%%%%%%%%
% Please check with the editor of your book whether he plans to
% include a "mutual" subject index - if so, please code your entries
% in the standard syntax. For your own purposes you may print your
% "personal" index by using the following commands:
%
%\clearpage
%\addcontentsline{toc}{section}{Index}
%\flushbottom
%\printindex
%%%%%%%%%%%%%%%%%%%%%%%%%%%%%%%%%%%%%%%%%%%%%%%%%%%%%%%%%%%%%%%%%%%%%

\end{document}